# Simple alternative to the Hardy-Ramanujan-Rademacher formula for p(N)

N. M. Chase, Ph.D.
School of Arts and Sciences
Massachusetts College of Pharmacy and Health Sciences
179 Longwood Avenue
Boston, MA  02115   U.S.A.

nchase@mcp.edu

MSC: 05A17 Enumerative combinatorics: Partitions of integers

Keywords: integer partitions, number theory

October 12,  2003

## Abstract

A recent paper examined the global structure of integer partitions sequences and, via combinatorial analysis using modular arithmetic, derived a closed form expression for a map from (N, M) to the set of all partitions of a positive integer N into exactly M positive integer summands. The output of the IPS map was a "matrix" having M columns and a number of rows equal to p[N, M], the number of partitions of N into M parts. The global structure of integer partition sequences (IPS) is that of a complex tree. In this paper, we examine the structure of the IPS tree and, by counting the number of directed paths through the tree, obtain a simple formula which gives, in closed form, the total number of partitions of N into exactly M parts. By summing over M, we obtain a transparent alternative to the Hardy-Ramanujan-Rademacher formula for p(N).

# 1. Introduction

A recent paper [1] examined the global structure of integer partitions sequences [2 - 6]. Modular arithmetic was used to derive, in closed form, a map from (N, M) to the set of all partitions of a positive integer N into exactly M positive integer summands without regard to order. The output of the IPS map was a "matrix" having M columns and a number of rows equal to p[N, M], the number of partitions of N into M parts. The derivation of the IPS matrix was based on combinatorial analysis of an isomorphic problem, that of apprehending the global structure of the sequence of all ordered placement of N indistinguishable balls into M non-empty distinguishable bins. [1] demonstrated that the global organization of integer partition sequences is that of a complex tree, each distinct path through tree corresponding to a distinct partition of the integer N.

In this paper, we extend the work in [1] and seek a closed form expression for p[N,M], the number of partitions of N into exactly M integer summands. To this end, we examine in greater detail the structure of the IPS tree.

The paper is organized as follows. In Section 2, we briefly review the derivation of the IPS matrix, thereby providing necessary background for Section 3. In Section 3 we examine the global structure of the IPS tree and, by counting the number of paths through the tree, obtain p[N,M] for cases in which $3 \leq M < \frac{N}{2}$ and $\frac{N}{2} \leq M \leq N-3$. In Section 4, we extend the results of Section 3 to include the simpler partition structures and obtain p[N, M] for $1 \leq M \leq N$. We then sum over M from 1 to N to obtain an alternative to the Hardy-Ramanujan-Rademacher formula for p(N) [2, 5, 6].

## 2. Structure of the IPS Matrix

The IPS map derived in [1] apprehends the global structure of integer partitions sequences and outputs a "matrix" consisting of the set of all partitions of a positive integer N into exactly M positive integer summands

$$\{s_0, s_1, \ldots, s_{M-1}\}, \tag{1}$$

where

$$\sum_{i=0}^{M-1} s_i = N \tag{2}$$

and

$$s_0 \geq s_1 \ldots \geq s_{M-1} \geq 1. \tag{3}$$

As already noted, the IPS matrix also gives the set of all distributions of N indistinguishable balls in M non-empty distinguishable bins, subject to the condition (3). Each row of the IPS matrix corresponds to a distinct distribution of balls of the form (1), where the integer $s_i$ gives the occupancy of the i'th bin.

In [1], the IPS matrix is derived by the following strategy. Start with the most condensed distribution of N balls and then redistribute the balls in all possible ways consistent with (3) until the distribution becomes as widely dispersed as possible. Tracking this sequence of redistributions is facilitated by defining the quantity k = N - M.

In the most "condensed" distribution of balls over the M non-empty bins, we clearly have

$$s_i = \begin{cases} k+1, & i = 0 \\ 1, & 1 \leq i \leq M-1 \end{cases} \tag{4}$$

We note, for future reference, that the zero'th bin has k "excess balls" above the singlet occupancy of all others.

The IPS matrix is defined piecewise, having distinct structures for $M < k$ and $M \geq k$.

If M is less than k (M less than N/2), then in the most widely "dispersed" state of the system, each of the M bins will have occupancy greater than one. On the other hand, if M is greater than or equal to k (M greater than or equal to N/2) then in the most dispersed distribution, k bins will have occupancy greater than 1 and M - k bins will have singlet occupancy.

## 2 A   Integer Partitions Matrix for $M < k$

It is straightforward to determine the set of all possible redistributions of the k excess balls initially in bin 0 into the other bins, subject to (3). Let the symbol j[i], $1 \leq i \leq M - 1$, represent the number of balls which have been shifted from the zeroth into the i'th bin. An individual row of the IPS matrix may then be written in a form which incorporates (2).

$$\left\{\left(k + 1 - \sum_{h=1}^{M-1} j[h]\right), (1 + j[1]), (1 + j[2]), \ldots, (1 + j[M-1])\right\}. \tag{5a}$$

That is,
$$S_i = \begin{cases} \left(k + 1 - \sum_{h=1}^{M-1} j[h]\right), & i = 0 \\ (1 + j[i]), & 1 \leq i \leq M - 1 \end{cases} \tag{5b}$$

Now ordering according to (3) requires

$$\left(k - \sum_{h=1}^{M-1} j[h]\right) \geq j[1] \geq j[2] \geq \ldots \geq j[M-1]. \tag{6}$$

Condition (6) can be satisfied if the full IPS matrix is generated by repeated, recursive evaluation of (5a) over the non-negative integers $j[i]$, as follows:

j[M - 1] is increased in unit steps from 0 up to a certain maximum value $J_{max}^{(-)}[M-1]$,

then j[M - 2] is increased in unit steps from j[M - 1] up to a certain maximum value $J_{max}^{(-)}[M-2]$,

then j[M - 3] is increased in unit steps from j[M - 2] up to a certain maximum value $J_{max}^{(-)}[M-3]$, . . . ,

then j[2] is increased in unit steps from j[3] up to a certain maximum value $J^{(-)}_{max}[2]$,

and finally j[1] is increased in unit steps from j[2] up to a certain maximum value $J^{(-)}_{max}[1]$;

the superscripts ( - ) denote partitions in which M < k.

The IPS matrix is thus constructed from a complex tree of j values; each row of the IPS matrix containing its own unique sequence of values for the integers j[i]. The primary combinatorial problem is that of finding a general expression for the quantities we have referred to as $J^{(-)}_{max}[a]$. Note that (2) and (6) together require that in general the values of the $J^{(-)}_{max}[a]$, a < M - 1, depend on input of the values of all j[h] from h = a + 1 to M - 1. A more descriptive notation for the quantity we have referred to as $J^{(-)}_{max}[a]$, a < M - 1, is thus $J^{(-)}_{max}[a, \{j[h]\}_{h=a+1}^{M-1}]$; the search for its general form is clearly best initiated at the "roots" of the IPS matrix tree of j values, at the values of integers j[M - 1], to which we now turn.

It is straightforward to show that the greatest value of $S_{M-1}$ is given by

$$S_{M-1} = \frac{M + k - Mod[M + k, M]}{M};  \qquad (11)$$

clearly then

$$J^{(-)}_{max}[M - 1] = \frac{M + k - Mod[M + k, M]}{M} - 1. \qquad (12a)$$

(12a) may be expressed more compactly as

$$J^{(-)}_{max}[M - 1] = \left\lfloor \frac{M + k}{M} \right\rfloor - 1, \qquad (12b)$$

where $\lfloor X \rfloor$ denotes the greatest integer less than or equal to X.

Now the integers j[M - 1] range from 0 up to the maximum value given by (12) and, for each value of j[M - 1], we have a sequence of possible j[M - 2] values, the greatest of which may be determined as follows. Given that some of the k excess originally in bin 0 have been shifted to bin M - 1, the remaining balls must ultimately be distributed as equally as possible between bin M - 2 and those to its left (distribution over M - 1 bins).
Thus,

$$J_{\max}^{(-)}[M-2, j[M-1]] = \frac{k - j[M-1] - Mod[k - j[M-1], M-1]}{M-1}. \quad (13)$$

Let us take this line of reasoning one step further. For each combination of values of j[M - 1] and j[M - 2], we have a sequence of possible values for j[M - 3], ranging from j[M - 2] up to a certain maximum value determined as follows. Given the number of the k excess balls originally in bin 0 which have been shifted into bins M - 1 and M - 2, the remaining ones must finally be distributed as equally as possible between bin M - 3 and those to *its* left (distribution over M - 2 bins).
Thus,

$$J_{\max}^{(-)}[M-3, \{j[M-1], j[M-2]\}] = \frac{k - j[M-1] - j[M-2] - Mod[k - j[M-1] - j[M-2], M-2]}{M-2}; \quad (14)$$

Leaping ahead to the integers j[2], similar reasoning implies

$$J_{\max}^{(-)}[2, \{j[h]\}_{h=3}^{M-1}] = \frac{k - \sum_{h=3}^{M-1} j[h] - Mod[k - \sum_{h=3}^{M-1} j[h], 3]}{3}; \quad (15)$$

given the redistribution of some of the k excess balls into bins to the right of bin 2, the remaining ones must ultimately be distributed as equally as possible between it (bin 2), bin 1 and bin 0 (3 bins altogether).

Similarly,

$$J_{\max}^{(-)}\left[1,\{j[h]\}_{h=2}^{M-1}\right] = \frac{k - \sum_{h=2}^{M-1} j[h] - \text{Mod}\left[k - \sum_{h=2}^{M-1} j[h], 2\right]}{2}, \tag{16}$$

and in general we have

$$J_{\max}^{(-)}\left[a,\{j[h]\}_{h=a+1}^{M-1}\right] = \frac{k - \sum_{h=a+1}^{M-1} j[h] - \text{Mod}\left[k - \sum_{h=a+1}^{M-1} j[h], a+1\right]}{a+1}, \quad a < M-1. \tag{17a}$$

(17a) may be written more compactly as

$$J_{\max}^{(-)}\left[a,\{j[h]\}_{h=a+1}^{M-1}\right] = \left\lfloor \frac{k - \sum_{h=a+1}^{M-1} j[h]}{M-1} \right\rfloor, \quad a < M-1. \tag{17b}$$

Thus, for partitions in which $M < k$ (and M > 2), the IPS matrix is given by

$$\left\{\left(k + 1 - \sum_{h=1}^{M-1} j[h]\right), (1 + j[1]), (1 + j[2]), \ldots, (1 + j[M-1])\right\}\Bigg|_{j[1]=j[2]}^{J_{\max}^{(-)}[1]} \Bigg|_{j[2]=j[3]}^{J_{\max}^{(-)}[2]} \cdots \Bigg|_{j[M-1]=0}^{J_{\max}^{(-)}[M-1]}, \tag{18}$$

where $J_{\max}^{(-)}[a]$, $1 \leq a \leq M-2$, is given by (17) and $J_{\max}^{(-)}[M-1]$ is given by (12);

the super and subscripted vertical bars in (18) are used to denote the repeated, recursive evaluations described in the remarks following equation (6).

For M = 1 and M = 2, we have the simpler partition structures

$$\begin{array}{ll} \{k+1\}, & M = 1 \\ \\ \left\{(k+1-j[1]), (1+j[1])\right\}\Big|_{j[1]=0}^{J_{\max}[M-1]}, & M = 2 \end{array} \tag{19}$$

## 2 B  Integer Partitions Matrix for $M \geq k$

Individual rows of the IPS matrix now have the form

$$\left\{\left(k+1-\sum_{h=1}^{k-1}j[h]\right),(1+j[1]),(1+j[2]),\ldots,(1+j[k-1]),1,\ldots\right\} . \qquad (20)$$

each row containing (M - k) summands equal to 1.  Comparing (20) with (5a), note that in (20) the upper limit of the sum is k - 1, since the k excess balls originally in bin 0 are now redistributed over at most k (rather than M) bins.

Ordering according to (3) now requires

$$\left(k-\sum_{h=1}^{k-1}j[h]\right) \geq j[1] \geq j[2] \geq \ldots \geq j[k-1] \qquad (21)$$

j[k - 1] takes on integer values of 0 or 1 only (c.f. the Hindenburg algorithm [ 2, 3]).  Maxima for the integers j[s], $1 \leq s \leq k-2$, may be obtained by the same combinatorial arguments which led to (17), again the only difference being that here redistributions are over k (rather than M) bins. We find

$$J_{\max}^{(+)}\left[a,\{j[h]\}_{h=a+1}^{k-1}\right] = \frac{k-\sum_{s=a+1}^{k-1}j[s] - Mod\left[k-\sum_{s=a+1}^{k-1}j[s], a+1\right]}{a+1}, \quad a < k-1, \qquad (22a)$$

where the superscript (+) denotes partitions in which $M \geq k$.

(22a) may be expressed more compactly as

$$J_{\max}^{(+)}\left[a,\{j[h]\}_{h=a+1}^{k-1}\right] = \left\lfloor \frac{k-\sum_{s=a+1}^{k-1}j[s]}{a+1} \right\rfloor, \quad a < k-1 . \qquad (22b)$$

Thus, for partitions in which $M \geq k$ (and k > 2), the IPS matrix is given by

$$\left\{\left(k+1-\sum_{h=1}^{k-1}j[h]\right),(1+j[1]),(1+j[2]),\ldots,(1+j[k-1]),1,\ldots\right\}\Bigg|_{j[1]=j[2]}^{J_{max}^{(+)}[1]}\Bigg|_{j[2]=j[3]}^{J_{max}^{(+)}[2]}\cdots\Bigg|_{j[k-2]=j[k-1]}^{J_{max}^{(+)}[k-2]}\Bigg|_{j[k-1]=0}^{1}, \quad (23)$$

where $J_{max}^{(+)}[a]$, a < k - 1, is an abbreviated notation for (22), and the super and subscripted vertical

bars in (23) indicate repeated recursive evaluation as follows:

the integer j[k - 1] is incremented from 0 to 1,

then j[k - 2] is increased from j[k - 1] up to a maximum $J_{max}^{(+)}[k-2]$, . . . ,

then j[2] is increased in unit steps from j[3] up to a maximum $J_{max}^{(+)}[2]$,

and finally j[1] is increased in unit steps from j[2] up to a maximum $J_{max}^{(+)}[1]$.

For k = 0, 1, and 2, we have the simpler partitions structures

$$\begin{array}{ll} \{1\}_{i=1}^{M}, & k=0 \\ \{2,\{1\}_{i=1}^{M-1}\}, & k=1 \\ \left\{(3-j[1]),(1+j[1]),\{1\}_{s=1}^{M-2}\right\}\Big|_{j[1]=0}^{1}, & k=2 \end{array} \quad (24)$$

## 3. Paths through the IPS Function's Tree of j Values

Let us now consider the global structure of the IPS matrix tree of j values. As the structure is fairly complex, it will be helpful to introduce some descriptive terminology while reviewing structures introduced in Section 2.

### 3 A   Tree Structure for $2 < M < k$

At the "roots" of the tree we find j[M - 1], which takes on integer values ranging from 0 up to

$$J^{(-)}_{max}[M-1] = \left\lfloor \frac{M+k}{M} \right\rfloor - 1 . \tag{12b}$$

As already noted, "emerging" from each of these j[M - 1] values (characterizing "nodes" at level M - 1 of the tree), we have branches to a set of possible integer values of j[M - 2], which range in unit steps from j[M - 1] up to

$$J^{(-)}_{max}[M-2, j[M-1]] = \left\lfloor \frac{k - j[M-1]}{M-1} \right\rfloor . \tag{25}$$

The greatest value of the integer j[M - 2], and thus the count of nodes at level M - 2 of the tree, depends on the particular j[M - 1] node value from which a set of j[M - 2] nodes emerges.

Emerging from each of the integer values of j[M - 2] (nodes at level M - 2), we have branches to a set of possible integer values of j[M - 3] (nodes at level M - 3), which range in unit steps from j[M - 2] up to

$$J^{(-)}_{max}[M-3, \{j[M-1], j[M-2]\}] = \left\lfloor \frac{k - j[M-1] - j[M-2]}{M-2} \right\rfloor . \tag{26}$$

The greatest value of j[M - 3], and thus the count of nodes at level M - 3, depends on the unique sequence of nodes ( j[M - 2] and j[M - 1] values) from which the set of j[M - 3] nodes emerges.

Now let us imagine exploring the structure of the entire tree by embarking from one of its root j[M -1] values and taking any one of the many paths along the tree's branches to a particular terminus (value of j[1]). (We imagine proceeding through a sequence of nodes j[s] with strictly decreasing "s"; we will refer to such paths through the tree as "directed paths". ) As we make our way along the branches we find that, from each of the nodes we encounter, branches emerge whose number depends on the values of all of the j's at the nodes we have traversed on our way to it (tracing all the way back to the root j[M - 1] value from which we embarked). Now to find the total number of partitions p[N,M] of an integer N into exactly M parts, we must find the total number of such directed paths through the tree, starting at each of the root j[M - 1] nodes, and passing to each of the possible nodes j[1]. Given the complex branching structure of the tree, such a counting process at first appears daunting. However, a counting strategy is easily built up by induction. Let us begin by counting the number of node values for j[M - 2].

For convenience, we will use the abbreviated notation $J_{max}^{(-)}[a]$, a < M - 1, to represent the quantity given in (17); $J_{max}^{(-)}[M-1]$ is given by (12).

Table 1 demonstrates that the number of nodes at level M - 2 of the tree is given by

$$n[M-2] = \sum_{j[M-1]=0}^{J_{max}^{(-)}[M-1]} \left(J_{max}^{(-)}[M-2] - j[M-1] + 1\right) = \sum_{j[M-1]=0}^{\left\lfloor \frac{M+k}{M} \right\rfloor - 1} \left(\left\lfloor \frac{k-j[M-1]}{M-1} \right\rfloor - j[M-1] + 1\right) \quad . \tag{27}$$

Table 2 demonstrates that the number of nodes at level M - 3 of the tree is given by

$$n[M-3] = \sum_{j[M-1]=0}^{J_{max}^{(-)}[M-1]} \sum_{j[M-2]=j[M-1]}^{J_{max}^{(-)}[M-2]} \left(J_{max}^{(-)}[M-3] - j[M-2] + 1\right). \tag{28a}$$

More explicitly,

$$n[M-3] = \sum_{j[M-1]=0}^{\left\lfloor \frac{M+k}{M} \right\rfloor - 1} \sum_{j[M-2]=j[M-1]}^{\left\lfloor \frac{k-j[M-1]}{M-1} \right\rfloor} \left(\left\lfloor \frac{k-j[M-1]-j[M-2]}{M-2} \right\rfloor - j[M-2] + 1\right). \tag{28b}$$

By a straightforward generalization of the structures evident in Tables 1 and 2, we may conclude that the number of nodes at level "a" of the tree is given by

$$n[a] = \sum_{j[M-1]=0}^{J_{max}^{(-)}[M-1]} \sum_{j[M-2]=j[M-1]}^{J_{max}^{(-)}[M-2]} \cdots \sum_{j[a+1]=j[a+2]}^{J_{max}^{(-)}[a+1]} \left(J_{max}^{(-)}[a] - j[a+1] + 1\right) \quad (29)$$

Clearly, the total number of directed paths through the tree is equal to the total number of nodes at level 1 of the tree. From (29), we find that for $2 < M < k$

$$n[1] = \sum_{j[M-1]=0}^{J_{max}^{(-)}[M-1]} \left( \sum_{j[M-2]=j[M-1]}^{J_{max}^{(-)}[M-2]} \cdots \sum_{j[3]=j[4]}^{J_{max}^{(-)}[3]} \sum_{j[2]=j[3]}^{J_{max}^{(-)}[2]} \left(J_{max}^{(-)}[1] - j[2] + 1\right) \right). \quad (30)$$

Now the total number of partitions $p_-[N, M]$ of a positive integer N into exactly M parts without regard to order is equal to the total number of paths through the tree (for M < k), given by (30).

Thus, for partitions in which $3 \leq M < N/2$,

$$p_-[N, M] = \sum_{j[M-1]=0}^{J_{max}^{(-)}[M-1]} \left( \sum_{j[M-2]=j[M-1]}^{J_{max}^{(-)}[M-2]} \cdots \sum_{j[3]=j[4]}^{J_{max}^{(-)}[3]} \sum_{j[2]=j[3]}^{J_{max}^{(-)}[2]} \left(J_{max}^{(-)}[1] - j[2] + 1\right) \right). \quad (31)$$

where $J_{max}^{(-)}[a]$, a < M - 1, is an abbreviated notation for

$$J_{max}^{(-)}\left[a, \{j[h]\}_{h=a+1}^{M-1}\right] = \left\lfloor \frac{N - M - \sum_{h=2}^{M-1} j[h]}{a+1} \right\rfloor, \quad a < M - 1 \quad (32)$$

and $J_{max}^{(-)}[M-1]$ is given by

$$J_{max}^{(-)}[M-1] = \left\lfloor \frac{N}{M} \right\rfloor - 1. \quad (33)$$

$p_-[N, M]$ for the simpler partitions in which M = 1 and M = 2 will be given in Section 4.

## 3 B   Tree Structure for $M \geq k > 2$

As already noted, only (k - 1) j's are required to account for all such partitions and $J_{max}^{(+)}[k-1] = 1$. Otherwise, arguments in Section 3A may be carried over without elaboration.

For convenience, we again adopt abbreviated notation and let $J_{max}^{(+)}[a]$, a < k - 1, denote the quantity given in (22).

Table 3 shows that the total number of nodes at level k - 2 of the tree is given by

$$n[k-2] = \sum_{j[k-1]=0}^{1} \left( J_{max}^{(+)}[k-2] - j[k-1] + 1 \right) = \sum_{j[k-1]=0}^{1} \left( \frac{k - j[k-1] - Mod\left[(k - j[k-1]), k-1\right]}{k-1} - j[k-1] + 1 \right) . \quad (34)$$

From Table 4 we conclude that the total of nodes at level k - 3 is given by

$$n[k-3] = \sum_{j[k-1]=0}^{1} \sum_{j[k-2]=j[k-1]}^{J_{max}^{(+)}[k-2]} \left( J_{max}^{(+)}[k-3] - j[k-2] + 1 \right). \quad (35a)$$

More explicitly,

$$n[k-3] = \sum_{j[k-1]=0}^{1} \sum_{j[k-2]=j[k-1]}^{\left\lfloor \frac{k-j[k-1]}{k-2} \right\rfloor} \left( \left\lfloor \frac{k - j[k-1] - j[k-2]}{k-2} \right\rfloor - j[k-2] + 1 \right). \quad (35b)$$

Straightforwardly generalizing the structures apparent in Tables 3 and 4, we obtain the number of nodes at level "a" of the tree

$$n[a] = \sum_{j[k-1]=0}^{1} \sum_{j[k-2]=j[k-1]}^{J_{max}^{(+)}[k-2]} \cdots \sum_{j[a+1]=j[a+2]}^{J_{max}^{(+)}[a+1]} \left( J_{max}^{(+)}[a] - j[a+1] + 1 \right). \quad (36)$$

Clearly, the total number of directed paths through the tree is equal to the total number of nodes at level 1 of the tree. From (36), we obtain

$$n[1] = \sum_{j[k-1]=0}^{1} \sum_{j[k-2]=j[k-1]}^{J_{max}^{(+)}[k-2]} \cdots \sum_{j[2]=j[3]}^{J_{max}^{(+)}[2]} \left( J_{max}^{(+)}[1] - j[2] + 1 \right). \qquad (37)$$

The total number of partitions $p_+[N, M]$ of a positive integer N into exactly M parts without regard to order is equal to the total number of directed paths through the tree, given by (37).

Thus, for partitions in which $\frac{N}{2} \leq M \leq N - 3$ ($M \geq k \geq 3$)

$$p_+[N, M] = \sum_{j[k-1]=0}^{1} \sum_{j[k-2]=j[k-1]}^{J_{max}^{(+)}[k-2]} \cdots \sum_{j[2]=j[3]}^{J_{max}^{(+)}[2]} \left( J_{max}^{(+)}[1] - j[2] + 1 \right). \qquad (38)$$

where $J_{max}^{(+)}[a]$, a < k - 1, is used to denote

$$J_{max}^{(+)}\left[a, \{j[h]\}_{h=2}^{N-M-1}\right] = \left\lfloor \frac{N - M - \sum_{h=2}^{N-M-1} j[h]}{a + 1} \right\rfloor, \quad a < N - M - 1. \qquad (39)$$

$p_+[N, M]$ for the simpler partitions in which M = N –2, N–1, N (k = 0, 1, 2) will be given in Section 4, to which we now turn.

## 4. Alternative to the Hardy-Ramanujan-Rademacher Formula for p(N)

Discussion in Section 3A excluded the simpler special cases of partitions in which M < 3; Section 3B omitted the simpler partitions in which k < 3. In this section, we include these special cases, without discussion as they are easily obtained from (19) and (24).

For partitions in which M < N/2, we may extend (31) to obtain

$$P_{-}[N,M] = \begin{cases} \sum_{j[M-1]=0}^{J_{max}^{(-)}[M-1]}\left(\sum_{j[M-2]=j[M-1]}^{J_{max}^{(-)}[M-2]} \cdots \sum_{j[2]=j[3]}^{J_{max}^{(-)}[2]}\left(J_{max}^{(-)}[1]-j[2]+1\right)\right), & 3 \leq M < N/2,\ N \geq 7 \\ \dfrac{N - Mod[N,2]}{2}, & M = 2 \\ 1, & M = 1 \end{cases} \quad (40)$$

where $J_{max}^{(-)}[a]$, a < M - 1, is an abbreviated notation for

$$J_{max}^{(-)}\left[a, \{j[h]\}_{h=a+1}^{M-1}\right] = \left\lfloor \dfrac{N - M - \sum_{s=a+1}^{M-1} j[s]}{a+1} \right\rfloor, \quad a < M - 1 \quad (41)$$

and

$$J_{max}^{(-)}[M-1] = \left\lfloor \dfrac{N}{M} \right\rfloor - 1 \quad . \quad (42)$$

Similarly, for partitions in which $M \geq N/2$, (38) may be extended to obtain

$$P_+[N,M] = \begin{cases} \sum_{j[N-M-1]=0}^{1} \left( \sum_{j[N-M-2]=j[N-M-1]}^{J_{max}^{(+)}[N-M-2]} \cdots \sum_{j[2]=j[3]}^{J_{max}^{(+)}[2]} \left( J_{max}^{(+)}[1] - j[2] + 1 \right) \right), & N/2 \leq M \leq N-3,\ N \geq 6 \\ 2, & M = N-2 \\ 1, & M = N-1 \\ 1, & N = M \end{cases}, \quad (43)$$

where

$$J_{max}^{(+)}\left[a, \{j[h]\}_{h=a+1}^{N-M-1}\right] = \left\lfloor \frac{N - M - \sum_{s=a+1}^{N-M-1} j[s]}{a+1} \right\rfloor, \quad a < N - M - 1. \quad (44)$$

Now the total number of partitions P[N] of a positive integer N is given by

$$\sum_{M=1}^{N} P[N, M]. \quad (45)$$

Since the IPS matrix is defined piecewise, (40) through (44) must be substituted into (45).

For N even, we obtain

$$P[N] = \sum_{M=1}^{\frac{N-Mod[N,2]}{2}-1} P_-[N, M] + \sum_{M=\frac{N-Mod[N,2]}{2}}^{N} P_+[N, M] ; \quad (46a)$$

for N odd,

$$P[N] = \sum_{M=1}^{\frac{N-Mod[N,2]}{2}} P_-[N, M] + \sum_{M=\frac{N-Mod[N,2]}{2}+1}^{N} P_+[N, M]. \quad (46b)$$

For even $N \geq 7$, we obtain

$$P[N] = 5 + \frac{N - Mod[N,2]}{2} + \sum_{M=3}^{\frac{N-Mod[N,2]}{2}-1} \left( \sum_{j[M-1]=0}^{J_{max}^{(-)}[M-1]} \left( \sum_{j[M-2]=j[M-1]}^{J_{max}^{(-)}[M-2]} \cdots \sum_{j[2]=j[3]}^{J_{max}^{(-)}[2]} \left( J_{max}^{(-)}[1] - j[2] + 1 \right) \right) \right)$$

$$+ \sum_{M=\frac{N-Mod[N,2]}{2}}^{N-3} \left( \sum_{j[N-M-1]=0}^{1} \left( \sum_{j[M-2]=j[M-1]}^{J_{max}^{(+)}[N-M-2]} \cdots \sum_{j[2]=j[3]}^{J_{max}^{(+)}[2]} \left( J_{max}^{(+)}[1] - j[2] + 1 \right) \right) \right) \quad ; \quad (47a)$$

for odd $N \geq 7$,

$$P[N] = 5 + \frac{N - Mod[N,2]}{2} + \sum_{M=3}^{\frac{N-Mod[N,2]}{2}} \left( \sum_{j[M-1]=0}^{J_{max}^{(-)}[M-1]} \left( \sum_{j[M-2]=j[M-1]}^{J_{max}^{(-)}[M-2]} \cdots \sum_{j[2]=j[3]}^{J_{max}^{(-)}[2]} \left( J_{max}^{(-)}[1] - j[2] + 1 \right) \right) \right)$$

$$+ \sum_{M=\frac{N-Mod[N,2]}{2}+1}^{N-3} \left( \sum_{j[N-M-1]=0}^{1} \left( \sum_{j[M-2]=j[M-1]}^{J_{max}^{(+)}[N-M-2]} \cdots \sum_{j[2]=j[3]}^{J_{max}^{(+)}[2]} \left( J_{max}^{(+)}[1] - j[2] + 1 \right) \right) \right) . \quad (47b)$$

Appendix 2 provides illustrative examples of calculations using (40) and (43).

# Appendix 1

Table 1: Counting the number n[M - 2] of values of the integer j[M - 2]. Superscripts ( - ) have been omitted from the table for convenience.

| Values of j[M - 1] | Values of j[M - 2] | Number of values of j[M - 2] |
|---|---|---|
| 0 | 0 | |
|   | 1 | |
|   | ... | Jmax[M - 2, j[M - 1]] + 1 |
|   | Jmax[M - 2, j[M - 1]] | |
| 1 | 1 | |
|   | 2 | Jmax[M - 2, j[M - 1]] |
|   | ... | |
|   | Jmax[M - 2, j[M - 1]] | |
| 2 | 2 | |
|   | 3 | Jmax[M - 2, j[M - 1]] - 1 |
|   | ... | |
|   | Jmax[M - 2, j[M - 1]] | |
| ... | ... | ... |
| Jmax[M - 1] | Jmax[M - 1] | |
|   | Jmax[M - 1] + 1 | Jmax[M - 2, j[M - 1]] - Jmax[M - 1] + 1 |
|   | ... | |
|   | Jmax[M - 2, j[M - 1]] | |

$$n[M-2] = \sum_{j[M-1]=0}^{J_{\max}^{(-)}[M-1]} \left( J_{\max}^{(-)}[M-2, j[M-1]] - j[M-1] + 1 \right) \quad \text{(A1)}$$

Table 2: Counting the number n[M - 3] of values of the integer j[M - 3], for j[M - 1] = 0. Superscripts ( - ) have been omitted from the table for convenience.

| Values of j[M - 2] | Values of j[M - 3] | Number of j[M - 3] values |
|---|---|---|
| 0 | 0 | Jmax[M - 3, {j[M - 2], j[M - 1]}] + 1 |
|   | 1 |   |
|   | ... |   |
|   | Jmax[M - 3, {j[M - 2], j[M - 1]}] |   |
| 1 | 1 | Jmax[M - 3, {j[M - 2], j[M - 1]}] |
|   | 2 |   |
|   | ... |   |
|   | Jmax[M - 3, {j[M - 2], j[M - 1]}] |   |
| 2 | 2 | Jmax[M - 3, {j[M - 2], j[M - 1]}] - 1 |
|   | 3 |   |
|   | ... |   |
|   | Jmax[M - 3, {j[M - 2], j[M - 1]}] |   |
| ... | ... | ... |
| Jmax[M - 2, j[M - 1]] | Jmax[M - 2, j[M - 1]] | Jmax[M - 3, {j[M - 2], j[M - 1]}] - Jmax[M - 2, j[M - 1]] + 1 |
|   | Jmax[M - 2, j[M - 1]] + 1 |   |
|   | ... |   |
|   | Jmax[M - 3, {j[M - 2], j[M - 1]}] |   |

$$n[M-3] = \sum_{j[M-1]=0}^{J_{max}^{(-)}[M-1]} \sum_{j[M-2]=j[M-1]}^{J_{max}^{(-)}[M-2,j[M-1]]} \left( J_{max}^{(-)}[M-3, \{j[M-2], j[M-1]\}] - j[M-2] + 1 \right) \quad (A2)$$

Table 3: Counting the number n[k - 2] of values of the integer j[k - 2]. Superscripts ( + ) have been omitted from the table for convenience.

| Values of j[k - 1] | Values of j[k - 2] | Number of values of j[k - 2] |
|---|---|---|
| 0 | 0 | |
| | 1 | Jmax[k - 2, j[k - 1]] + 1 |
| | ... | |
| | Jmax[k - 2, j[k - 1]] | |
| 1 | 1 | |
| | 2 | Jmax[ k - 2, j[k - 1]] |
| | ... | |
| | Jmax[ k - 2, j[k - 1]] | |

$$n[k-2] = \sum_{j[k-1]=0}^{1} \left( J_{\max}^{(+)}[k-2] - j[k-1] + 1 \right) \quad \text{(A3)}$$

Table 4: Counting the number n[k - 3] of values of the integer j[k - 3] , for j[k - 1] = 0. Superscripts ( + ) have been omitted from the table for convenience.

| j[k - 2] | j[k - 3] | Number of j[k - 3} values |
|---|---|---|
| 0 | 0 | |
| | 1 | Jmax[k - 3, {j[k - 2], j[k - 1]}] + 1 |
| | ... | |
| | Jmax[k - 3, {j[k - 2], j[k - 1]}] | |
| 1 | 1 | |
| | 2 | Jmax[k - 3, {j[k - 2], j[k - 1]}] |
| | ... | |
| | Jmax[k - 3, {j[k - 2], j[k - 1]}] | |
| 2 | 2 | |
| | 3 | Jmax[k - 3, {j[k - 2], j[k - 1]}] - 1 |
| | ... | |
| | Jmax[k - 3, {j[k - 2], j[k - 1]}] | |
| ... | ... | |
| Jmax[k - 2, j[k - 1]] | Jmax[k - 2, j[k - 1]] | |
| | Jmax[k - 2, j[k - 1]] + 1 | Jmax[k - 3, {j[k - 2], j[k - 1]}] - Jmax[k - 2, j[k - 1]] + 1 |
| | ... | |
| | Jmax[k - 3, {j[k - 2], j[k - 1]}] | |

$$n[k-3] = \sum_{j[k-1]=0}^{1} \sum_{j[k-2]=j[k-1]}^{J_{\max}^{(+)}[k-2]} \left( J_{\max}^{(+)}[k-3, \{j[k-2], j[k-1]\}] - j[k-2] + 1 \right) \quad (A4)$$

# Appendix 2
Sample Calculations Using Equations (40) and (43)

N = 6, M = 3, k = 3; (43) implies

$$P[6,3] = \sum_{j[2]=0}^{1} \left( \frac{3 - j[2] - Mod[3 - j[2],2]}{2} - j[2] + 1 \right) = 2 + 1 = 3.$$

N = 7, M = 4, k = 3; (43) implies

$$P[7,4] = \sum_{j[2]=0}^{1} \left( \frac{3 - j[2] - Mod[3 - j[2],2]}{2} - j[2] + 1 \right) = 2 + 1 = 3.$$

N = 7, M = 3, k = 4; (40) implies

$$P[7,3] = \sum_{j[2]=0}^{1} \left( \frac{4 - j[2] - Mod[4 - j[2],2]}{2} - j[2] + 1 \right) = 3 + 1 = 4.$$

N = 8, M = 5, k = 3; (43) implies

$$P[8,5] = \sum_{j[2]=0}^{1} \left( \frac{3 - j[2] - Mod[3 - j[2],2]}{2} - j[2] + 1 \right) = 2 + 1 = 3.$$

N = 8, M = 4, k = 4; (43) implies

$$P[8,4] = \sum_{j[3]=0}^{1} \sum_{j[2]=j[3]}^{1} \left( \frac{4 - j[3] - j[2] - Mod[4 - j[3] - j[2],2]}{2} - j[2] + 1 \right) = 3 + 1 + 1 = 5.$$

N = 8, M = 3, k = 5; (40) implies

$$P[8,3] = \sum_{j[2]=0}^{1} \left( \frac{5 - j[2] - Mod[5 - j[2],2]}{2} - j[2] + 1 \right) = 3 + 2 = 5.$$

N = 9, M = 6, k = 3; (43) implies

$$P[9,6] = \sum_{j[2]=0}^{1} \left( \frac{3 - j[2] - Mod[3 - j[2],2]}{2} - j[2] + 1 \right) = 2 + 1 = 3.$$

N = 9, M = 5, k = 4;  (43) implies

$$P[9,5] = \sum_{j[3]=0}^{1} \sum_{j[2]=j[3]}^{1} \left( \frac{4 - j[2] - j[3] - Mod[4 - j[2] - j[3], 2]}{2} - j[2] + 1 \right) = 3 + 1 + 1 = 5.$$

N = 9, M = 4, k = 5;  (40) implies

$$P[9,4] = \sum_{j[3]=0}^{1} \sum_{j[2]=j[3]}^{1} \left( \frac{5 - j[2] - j[3] - Mod[5 - j[2] - j[3], 2]}{2} - j[2] + 1 \right) = 3 + 2 + 1 = 6.$$

N = 9, M = 3, k = 6;  (40) implies

$$P[9,3] = \sum_{j[2]=0}^{2} \left( \frac{6 - j[2] - Mod[6 - j[2], 2]}{2} - j[2] + 1 \right) = 4 + 2 + 1 = 7.$$

N = 10, M = 7, k = 3;  (43) implies

$$P[10,7] = \sum_{j[2]=0}^{1} \left( \frac{3 - j[2] - Mod[3 - j[2], 2]}{2} - j[2] + 1 \right) = 2 + 1 = 3.$$

N = 10, M = 6, k = 4;  (43) implies

$$P[10,6] = \sum_{j[3]=0}^{1} \sum_{j[2]=j[3]}^{1} \left( \frac{4 - j[2] - j[3] - Mod[4 - j[2] - j[3], 2]}{2} - j[2] + 1 \right) = 3 + 1 + 1 = 5$$

N = 10, M = 5, k = 5;  (43) implies

$$P[10,5] = \sum_{j[4]=0}^{1} \sum_{j[3]=j[4]}^{1} \sum_{j[2]=j[3]}^{1} \left( \frac{5 - j[2] - j[3] - j[4] - Mod[5 - j[2] - j[3] - j[4], 2]}{2} - j[2] + 1 \right) = 3 + 2 + 1 + 1 = 7$$

N = 10, M = 4, k = 6;  (40) implies

$$P[10,4] = \sum_{j[3]=0}^{0} \sum_{j[2]=j[3]}^{2} \left( \frac{6 - j[2] - j[3] - Mod[6 - j[2] - j[3], 2]}{2} - j[2] + 1 \right)$$
$$+ \sum_{j[3]=1}^{1} \sum_{j[2]=j[3]}^{1} \left( \frac{6 - j[2] - j[3] - Mod[6 - j[2] - j[3], 2]}{2} - j[2] + 1 \right) = 4 + 2 + 1 + 2 = 9$$

N = 10, M = 3, k = 7;  (40) implies

$$P[10,3] = \sum_{j[2]=0}^{2} \left( \frac{7 - j[2] - Mod[7 - j[2], 2]}{2} - j[2] + 1 \right) = 4 + 3 + 1 = 8 \cdot$$